\documentclass[fleqn]{2017SCGE}
\usepackage{multirow}
\newcommand\aj{AJ}                   % Astronomical Journal (the)
\newcommand\apj{ApJ}                 % Astrophysical Journal
\newcommand\mnras{MNRAS}             % Monthly Notices of the Royal Astronomical Society
\newcommand\nat{Nature}              % Nature
\newcommand\pasa{Publ. Astron. Soc. Australia}  % Publications of the Astronomical Society of Australia
\newcommand{\sech}{\,\mathrm{sech}}
\newcommand{\degree}{^{\circ}}

\begin{document}

\ensubject{subject}

%%%%%%%%%%%%%%%%%%%%%%%%%%%%%%%%%%%%%%%%%%%%%%%%%%%%%%%
%%% Authors do not modify the information below
%%% ????????????????
%%% ??????????, ????????????{}, ???????????????????
%Letter to the Editor??Article%??????
\ArticleType{Article}%??Article
\SpecialTopic{SPECIAL TOPIC: FAST special issue}%???????
\Year{2018}
\Month{February}
\Vol{60}
\No{1}
\DOI{xxx}
\ArtNo{000000}
\ReceiveDate{February xx, 2018}
\AcceptDate{xx xx, 2018}
%\OnlineDate{January 1, 2016}
%%%%%%%%%%%%%%%%%%%%%%%%%%%%%%%%%%%%%%%%%%%%%%%%%%%%%%%

%%% title: ????
%%%   \title{title}{title for citation}
\title{Study of Three Rotating Radio Transients with FAST}

%%% Corresponding author: ???????
%%%   \author[number]{Full name}{{email@xxx.com}}
%%% General author: ???????
%%%   \author[number]{Full name}{}
\author[1,2]{Jiguang LU}{lujig@nao.cas.cn}%
\author[1,2]{Bo PENG}{pb@nao.cas.cn}
\author[3,1]{Kuo LIU}{}
\author[1]{\\Peng JIANG}{}
\author[1]{Youling YUE}{}
\author[1]{Meng YU}{}
\author[1,4]{Ye-Zhao YU}{}
\author[1]{Feifei KOU}{}
\author[1,4]{Lin WANG}{}
\author[1]{\\the FAST Collaboration}{}

%%% Author information for page head. ?��?��????????
%%% ??????????????, ??????????author???
\AuthorMark{Lu J G}%\authorcr????????

%%% Authors for citation. ????????��????????
%%% ??????????????, ??????????author???
\AuthorCitation{Lu J G, Peng B, Liu K et al.}

%%% Address. ???
%%%   \address[number]{Address, City {\rm Postcode}, Country}
\address[1]{CAS Key Laboratory of FAST, National Astronomical Observatories, Chinese Academy of Sciences, Beijing 100101, China}
\address[2]{Guizhou Radio Astronomy Observatory, Guizhou, China}
\address[3]{Max-Planck-Institut f$\ddot{u}$r Radioastronomie, Auf dem H$\ddot{u}$gel 69, D-53121 Bonn, Germany}
\address[4]{College of Astronomy and Space Sciences, University of Chinese Academy of Sciences, Beijing 100049, China}

%\contributions{}%????????

%%% Abstract. ??
\abstract{%
Rotating radio transients (RRATs) are peculiar astronomical objects whose emission mechanism remains under investigation.
In this paper, we present observations of three RRATs, J1538+2345, J1854+0306 and J1913+1330, observed with
the Five-hundred-meter Aperture Spherical radio Telescope (FAST).
Specifically, we analyze the mean pulse profiles and temporal flux density evolutions of the RRATs.
Owing to the high sensitivity of FAST, the derived burst rates of the three RRATs
are higher than those in previous reports.
RRAT J1854+0306 exhibited a time-dynamic mean pulse profile,
whereas RRAT J1913+1330 showed distinct radiation and nulling segments on its pulse intensity trains.
The mean pulse profile variation with frequency is also studied for RRAT J1538+2345 and RRAT J1913+1330,
and the profiles at different frequencies could be well fitted with a cone-core model and a conal-beam model, respectively.
}%ÕªÒª

%%% Keywords. ?????

%Equations of state of neutron-star matter; Acoustic signal processing; Control theory

\maketitle

%\tableofcontents%?????

%%%%%%%%%%%%%%%%%%%%%%%%%%%%%%%%%%%%%%%%%%%%%%%%%%%%%%%
%%% The main text. ???????
%???????????????????\cref{fig1}
%\twocolumn\onecolumn
%%%%%%%%%%%%%%%%%%%%%%%%%%%%%%%%%%%%%%%%%%%%%%%%%%%%%%%
\begin{multicols}{2}
\section{Introduction}\label{sect:intro}
\label{intro}
Prior to their discovery in 2006, rotating radio transients (RRATs) were an unknown population of bursting neutron stars~\cite{mcla06}.
RRATs emit bursts at radio frequencies with apparent randomness but with an underlying periodicity.
Despite that RRATs have been believed to be neutron stars mostly for their characteristics of pulse emission and periodicity \cite{kean11}, they have been classified as a group of radio sources based on their peculiar radiation phenomenon rather than their intrinsic properties and their nature of emission is still to be fully understood.
\Authorfootnote
\noindent
Some studies have identified links between RRATs and the other pulsar populations.
For instance,  Bhattacharyya et al. (2018) \cite{bhat18} found a post-glitch over-recovery
in the frequency derivative for
RRAT J1819$-$1458 which typifies a magnetar.
The surface magnetic field ($\sim4.94\times10^{13}$\,G)
also implies a magnetar nature of RRAT J1819$-$1458.
If RRAT J1819-1458 is indeed a magnetar, it will provide a useful reference for studying the connection between radio pulsars and the magnetar population \cite{cami07}.
Weltevrede et al. (2006) \cite{welt06} showed that bursts from PSR B0656+14 have the same characteristics as RRATs; thus, this object is possibly classifiable as a near RRAT.
Additionally, if considered as pulsars, RRATs exhibit an extreme nulling phenomenon, staying in the quiet state for most of the time.
Wang et al. (2007) \cite{wang07} speculated that large null fractions in nulling pulsars are more related to large characteristic age than long period. However, the ages of most RRATs are not yet determined.

For a detailed classification of RRATs based on their intrinsic properties, we must further probe the connection between RRATs and other neutron star populations.
For this purpose, more RRAT bursts and data with higher signal-to-noise ratio need to be collected and studied.
Notably, the burst rate of a RRAT may relate to the sensitivity of the telescope~\cite{mcla09}; a more sensitive telescope will detect
more bursts from a RRAT.
Therefore, the Five-hundred-meter Aperture Spherical radio Telescope (FAST),
the largest filled-aperture single-dish radio telescope at present~\cite{peng00a,peng00b}, is ideal for studying RRATs.

The remainder of this paper is structured as follows.
Section~\ref{sect:obs} details our observation and data flow,
and Section~\ref{sec:rrats} presents the results of the data analysis.
Our findings are discussed in Section~\ref{sect:discussion} and a brief summary is given in Section~\ref{sect:summary}.

\section{Observation and Data Reduction}
\label{sect:obs}

FAST is located in Guizhou, China, at longitude 106.9$\degree$\,E and latitude 25.7$\degree$\,N, and is currently in its commissioning phase.
The aperture of the telescope is 500\,m, of which 300\,m is effective.
In total, we obtained eleven observations on three RRATs, J1538+2345, J1854+0306 and J1913+1330 selected from the RRATalog\footnote{see http://astro.phys.wvu.edu/rratalog/} based on their coordinates and the sky coverage of FAST.
Each 30\,min observation was made using either an ultra-wideband receiver covering 270$-$1620\,MHz, or a 19-beam receiver covering 1000$-$1500\,MHz.
Details of the observations are provided in Table~\ref{tab1}.
%
%The system temperature is about 50\,K.
%
The data from the ultra-wideband receiver were filtered into two bands, a low-frequency band covering 270$-$800\,MHz (the P-band)
and a high-frequency band covering 1200$-$1620\,MHz (the L-band).
The data in each band were captured at the Nyquist rate with a digital backend based on Reconfigurable Open Architecture Computing Hardware generation 2 (ROACH 2)\footnote{https://casper.berkeley.edu/} \cite{jian19}.
The captured data were pre-processed and stored in PSRFITS format search mode~\cite{hota04}, with a time resolution of 0.2\,ms and frequency resolution of 0.25\,MHz
(P-band and L-band data were recorded simultaneously).
The data from the central beam of the 19-beam receiver were recorded in the same format but with finer resolution (time and frequency resolutions of 49.152\,$\mathrm{\mu}$s and 0.122\,MHz, respectively).
Later, the data of RRAT J1538+2345 and RRAT J1854+0306 were incoherently de-dispersed with dispersion measures (DMs) of 14.909\,$\mathrm{cm^{-3}\,pc}$~\cite{kara15} and 192.4\,$\mathrm{cm^{-3}\,pc}$~\cite{kean11}, respectively.
In each observation of RRAT J1913+1330, the DM was determined from the P-band FAST data, and these DM values are reported in Section~\ref{J1913}.
Next the de-dispersed data were folded with the Chebyshev predictor calculated by
TEMPO2~\cite{hobb06,edwa06} with the ephemeris provided by PSRCAT\footnote{see
http://www.atnf.csiro.au/research/pulsar/psrcat/} (version 1.56,~\cite{manc05}).
The frequency channels with distinct radio interference were then removed manually.
To study the evolution of radiation at different frequencies, the data of the 19-beam and ultra-wideband receivers were divided into narrower bands of bandwidth 100\,MHz and 50\,MHz, respectively.

\begin{table}[H]
\centering
\footnotesize
\begin{threeparttable}\caption{Observing information.}\label{tab1}
%\doublerulesep 0.1pt \tabcolsep 13pt %space between two columns. ????????��???
\begin{tabular}{l|cccc}
\toprule
  Source & Obs Date  & Obs Band  & Total & Burst\\
  & (MJD) & (MHz) & Pulses & Pulses \\\hline
  \multirow{4}[0]*{RRAT J1538+2345} & 58338 & 1000$-$1500 & 521 & 183 \\
  & 58359 & 1000$-$1500 & 521 & 179 \\
  & 58395 & 1000$-$1500 & 521 & 108 \\
  & 58407 & 1000$-$1500 & 521 & 134 \\\hline
\multirow{2}[0]*{RRAT J1854+0306}  & 58391 & 1000$-$1500 & 394 & 46\\
  & 58395 & 1000$-$1500 & 394 & 56 \\\hline
\multirow{5}[0]*{RRAT J1913$+$1330} & 58085 & 270$-$800, 1200$-$1620 & 1975\tnote{1)} & 27 \\
  & 58140 & 270$-$800, 1200$-$1620 & 1977 & 0 \\
  & 58144 & 270$-$800, 1200$-$1620 & 1891 & 24 \\
  & 58149 & 270$-$800, 1200$-$1620 & 1777 & 167 \\
  & 58179 & 270$-$800, 1200$-$1620 & 1919 & 169 \\
\bottomrule
\end{tabular}
\begin{tablenotes}
\item[1)] the cycle number and burst number of RRAT J1913+1330 were derived from the P-band data.
\end{tablenotes}
\end{threeparttable}
\end{table}

\section{Radiation from the three RRATs}
\label{sec:rrats}

\subsection{RRAT J1538+2345}

\begin{figure}[H]
  \centering
   \includegraphics[width=0.9\linewidth]{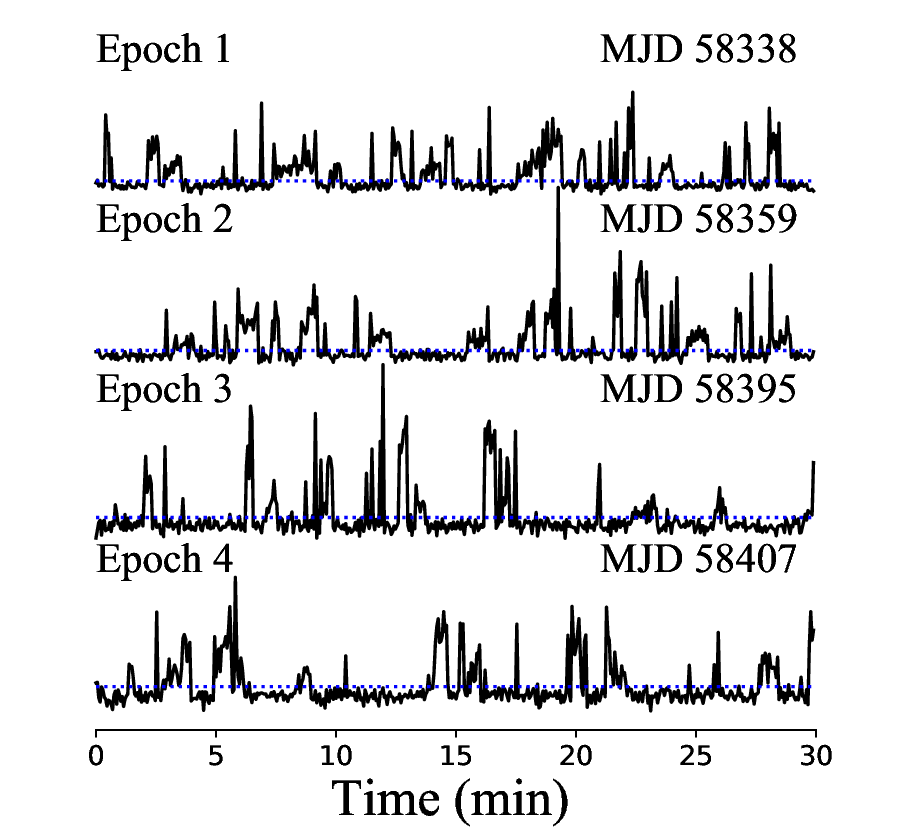}
   \caption{{\small Intensity time series of single pulse detections from RRAT J1538+2345.
   Each peak represents the intensity of one pulse cycle.
   The blue dotted line indicates 10 times the rms of the off-pulse region.}}
   \label{fig1}
\end{figure}

\begin{figure}[H]
  \centering
   \includegraphics[width=0.9\linewidth]{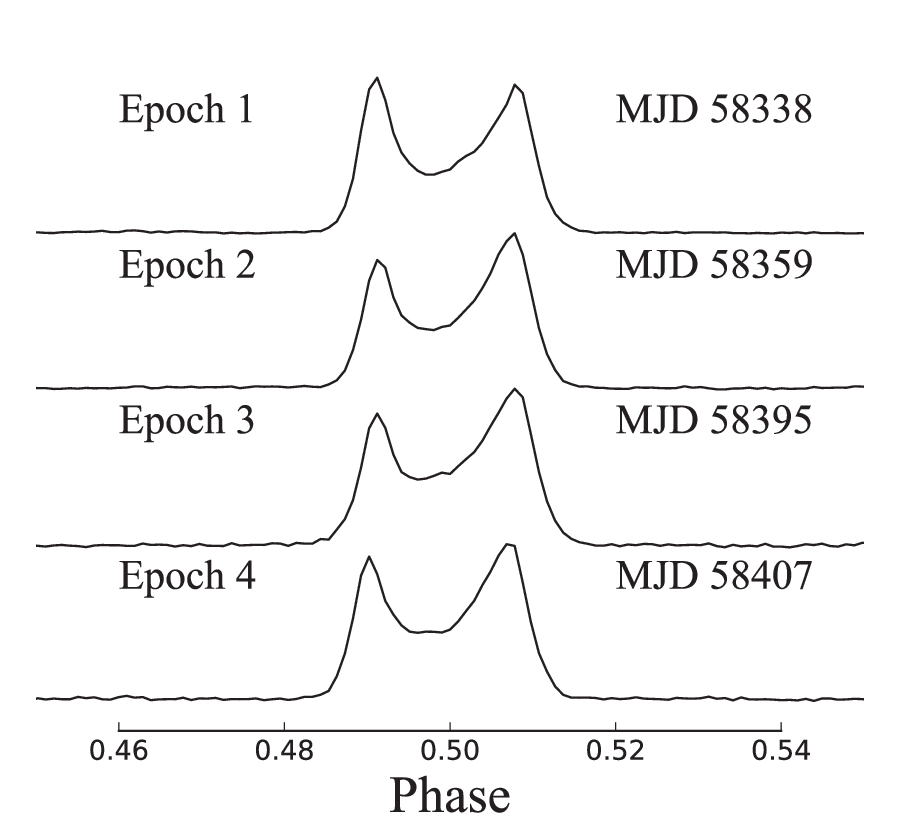}
   \caption{{\small Mean pulse profiles of RRAT J1538+2345 during each observation epoch.}}
   \label{fig2}
\end{figure}

RRAT J1538+2345, with a period of $\sim$3.45\,s~\cite{kean11}, was discovered in the Green Bank Telescope (GBT) 350 MHz Drift Survey.
In a 77 minute LOFAR observation with central frequency and bandwidth of 150\,MHz and 80\,MHz respectively,
the burst rate of RRAT J1538+2345 was determined as 66$\pm$7\,hr$^{-1}$~\cite{kara15} (see Table~\ref{tab2}).
Figure~\ref{fig1} shows the intensity time series of individual pulses (phases 0.47$-$0.53 in Figure~\ref{fig2}) during each observation epoch of FAST.
The detection threshold (blue dotted line in Figure~\ref{fig1}) was taken as 10 times the rms of the off-pulse region.
Most of the burst cycles of RRAT J1538+2345 occurred in continuous trains composed of tens of pulse cycles.
This phenomenon may relate to the intrinsic radiation mechanism of this pulsar.
The statistics of the detected bursts are given in Table~\ref{tab1}.
Over the total observation time of $\sim$120 minutes, the burst rate of RRAT J1538+2345 can be estimated
as $302\pm$12\,hr$^{-1}$, much larger than that determined in Karako-Argaman et al. (2015), 66$\pm$7\,hr$^{-1}$~\cite{kara15}.

\begin{figure}[H]
  \centering
   \includegraphics[width=0.9\linewidth]{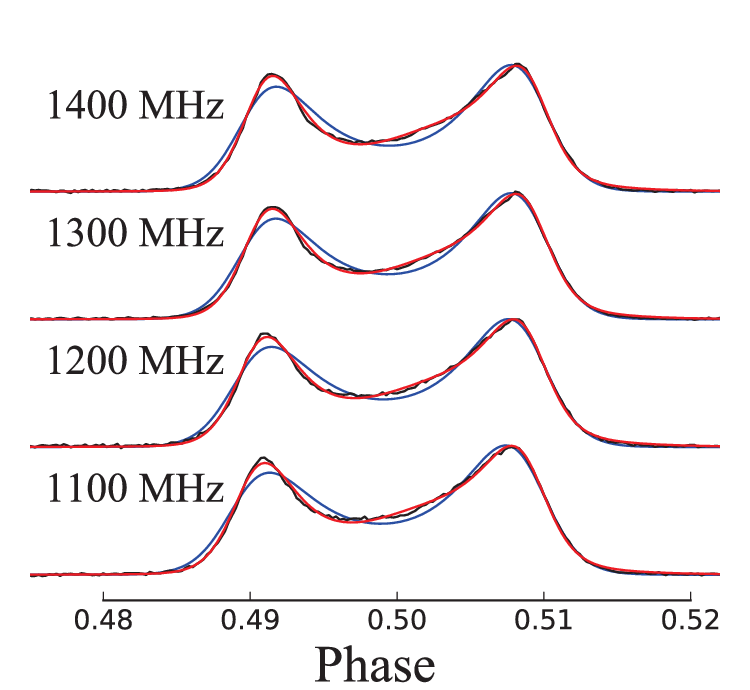}
   \caption{{\small Multi-frequency profiles of RRAT J1538+2345 (black curves).
   The blue and red curves are fitted by Equation~\ref{conalfunction} and Equation~\ref{conalfunction1}, respectively.}}
   \label{fig3}
\end{figure}

In each epoch, the bursts of RRAT J1538+2345 were selected and integrated to form a mean pulse profile, and the results are shown in Figure~\ref{fig2}.
Note that the mean pulse profile of RRAT J1538+2345 did not change significantly from epoch to epoch. This indicates that the radio emission from the magnetosphere of this pulsar is overall stable on a timescale of days.
Figure~\ref{fig3} shows the mean pulse profiles at different frequencies, and the profile variations are insignificant across our observing bands.
The mean pulse profile has two peaks separated by a bridge component, which characterizes a not-well-resolved conal-double profile~\cite{rank83} as the mean pulse profile of PSR B1133+16.
We then attempted to fit the profiles by the modelling approach of Lu et al. (2016) \cite{lu16a},
\begin{equation}
F_{\mathrm{conal~beam}}=
F_{\mathrm{sech~square}}(\theta_{\mu},A,\sigma,\theta_{\mu0})\times[1+k\cdot(\phi-\phi_0)],
\label{conalfunction}
\end{equation}
where $\phi$ is the pulse phase; $A$, $\phi_{0}$ and $\sigma$ are parameters describing the shape of the radiation cone;
$F_{\mathrm{sech~square}}$ is defined as follows (which has been proved to be a good function to describe the shape of pulse peak in Lu et al. (2016)~\cite{lu16a}),
\begin{equation}
F_{\mathrm{sech~square}}(\phi,A,\sigma,\phi_{0})=
A\sech^2\left(\sqrt{\frac{2}{\pi}}\frac{\phi-\phi_{0}}{\sigma}\right),
\label{sechfunction}
\end{equation}
where $\theta_{\mu0}$ is the angular radius of the conal beam, and the angular distance $\theta_{\mu}$ between the magnetic axis and the radiation direction is calculated as follows,
\begin{equation}
\cos\theta_{\mu}=\cos\alpha\cos(\alpha+\beta)+\sin\alpha\sin(\alpha+\beta)\cos(\phi-\phi_0),
\label{geometry}
\end{equation}
where $\alpha$ and $\beta$ are the magnetic inclination angle and impact angle respectively.
The profiles of each individual frequency band were fitted by the Levenberg--Marquardt algorithm (also adopted in the remainder of this paper), and the fitting results are shown in Figure~\ref{fig3} as blue curves.
As the profiles were poorly fitted by the conal-beam model alone, we added an additional component to the conal-beam function as follows,
\begin{equation}
F_{\mathrm{conal~core}}=F_{\mathrm{conal~beam}}+F_{\mathrm{sech~square}}(\phi,A,\sigma,\phi_{0}).
\label{conalfunction1}
\end{equation}
This model fitted the data significantly better than the conal-beam model, and the results are shown in Figure~\ref{fig3} as red curves.
With this fit, the best-estimated inclination angle and impact angle were determined as $\alpha=33.0\degree,~\beta=14.1\degree$, respectively.

\subsection{RRAT J1854+0306}

\begin{figure}[H]
  \centering
   \includegraphics[width=0.9\linewidth]{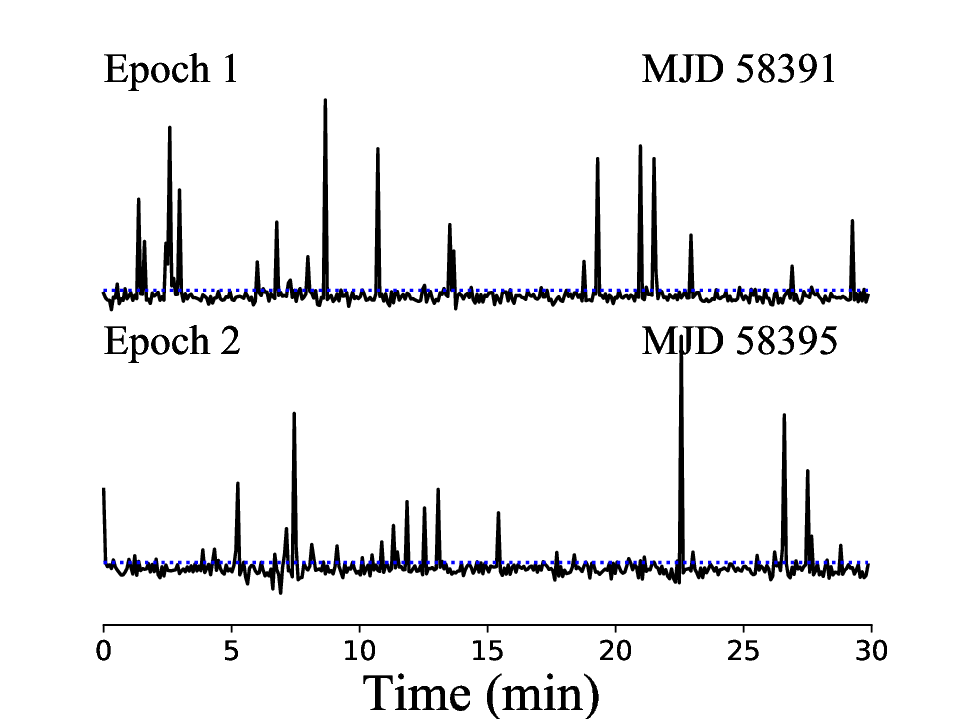}
   \caption{{\small As Figure~\ref{fig1}, but for RRAT J1854+0306.}}
   \label{fig4}
\end{figure}

RRAT J1854+0306, with a period of $\sim$4.56\,s~\cite{kean11}, was discovered in the Arecibo Pulsar ALFA survey~\cite{dene09}.
Table~\ref{tab2} lists the burst rate calculated by Keane et al. (2011) \cite{kean11}, 8.9\,hr$^{-1}$,
which is based on the L-band data with a central frequency of 1374\,MHz and a bandwidth of 288\,MHz observed with the Parkes Radio Telescope.
This value is much lower than that obtained in Deneva et al. (2019) \cite{dene09}, 84\,hr$^{-1}$, from the data with 1440\,MHz central frequency and 100\,MHz bandwidth
observed with the Arecibo Observatory.
Figure~\ref{fig4} shows the intensity time series of the individual pulses in our FAST observations, from which the burst rate was estimated as 102$\pm$10\,hr$^{-1}$.
This value is generally consistent with the result obtained in Deneva et al. (2019) \cite{dene09}.

\begin{figure}[H]
  \centering
   \includegraphics[width=0.9\linewidth]{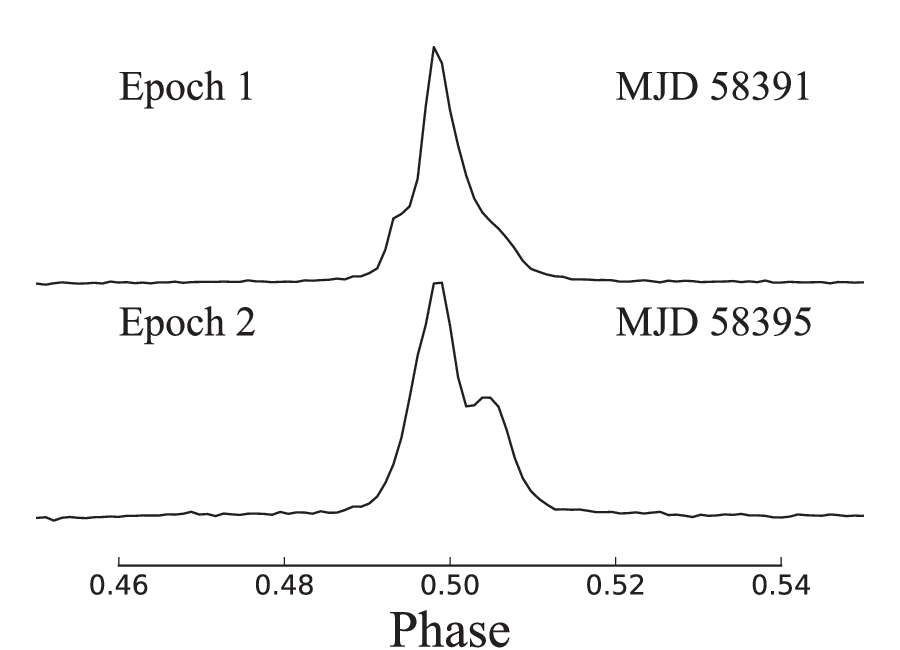}
   \caption{{\small As Figure~\ref{fig2}, but plotted for RRAT J1854+0306.}}
   \label{fig5}
\end{figure}

The burst pulses of RRAT J1854+0306 from each epoch were also extracted and integrated to form a mean pulse profile which is shown in Figure~\ref{fig5}.
Apparently, the mean profiles from the two epochs are quite different, indicating an intra-day variation of the magnetosphere of pulsar.
However, we cannot rule out that the number of pulses integrated ($\simeq50$) is insufficient to form a stable mean profile.

\subsection{RRAT J1913+1330}
\label{J1913}

\begin{figure*}
  \centering
   \includegraphics[width=0.45\linewidth]{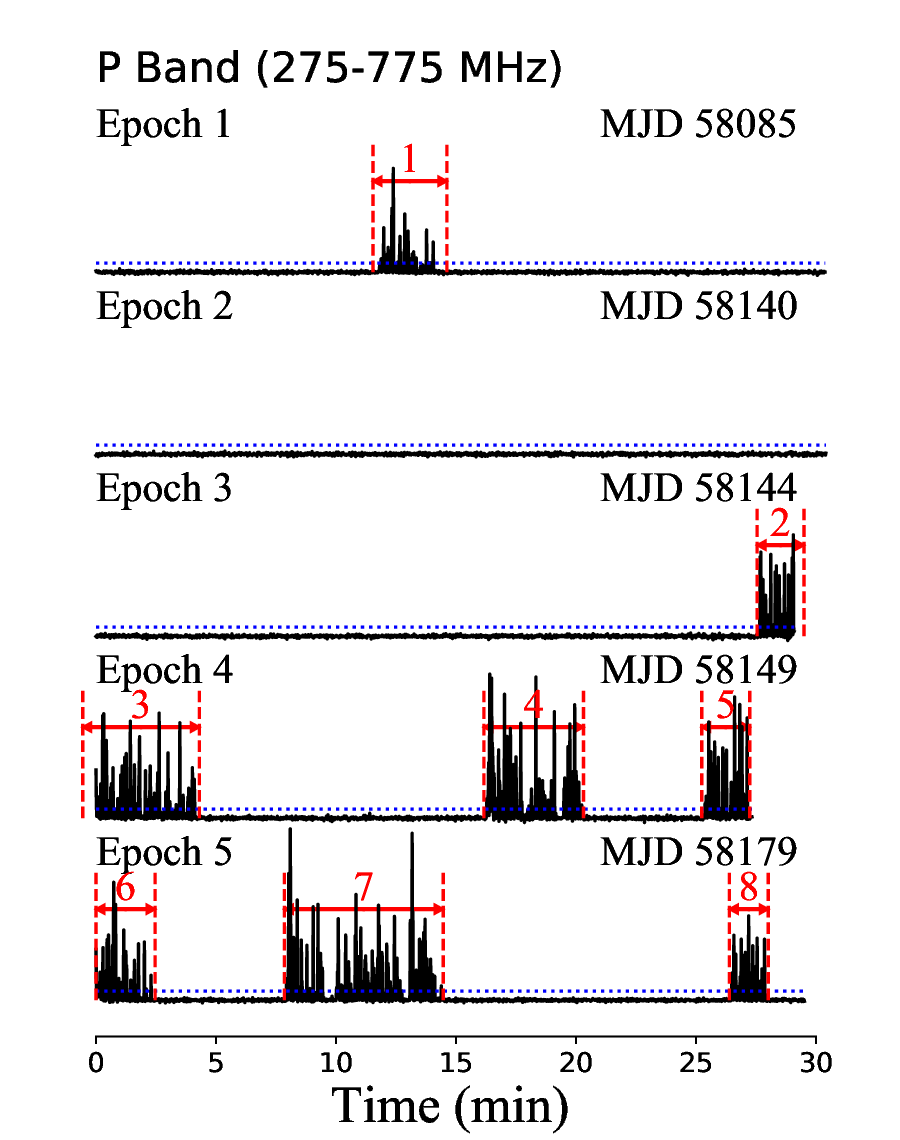}
   \includegraphics[width=0.45\linewidth]{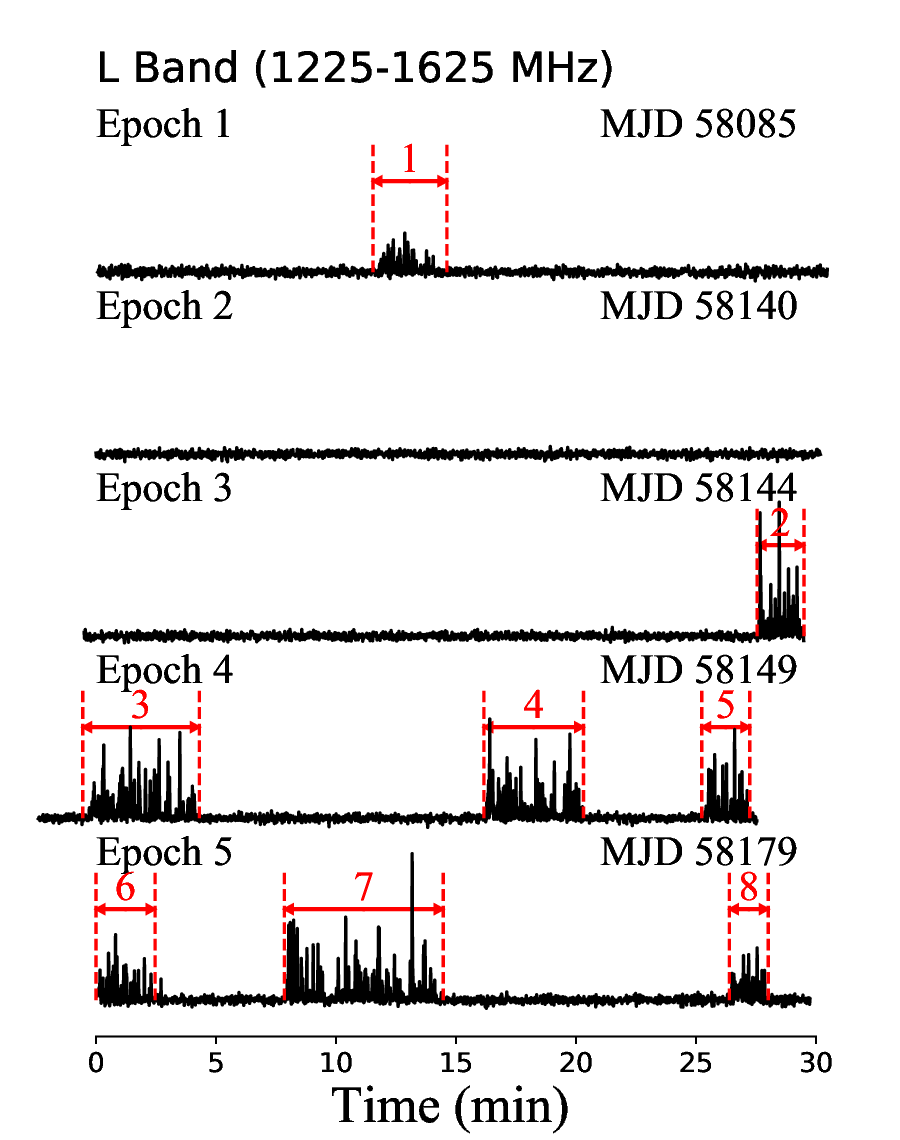}
   \caption{{\small Fluxes of pulse-on phase intervals of RRAT J1913+1330 at different frequency bands.
   The blue dotted lines in left panel indicate 10 times the rms of the pulse-off phase intervals.
   The radiation segments are numbered successively.}}
   \label{fig6}
\end{figure*}

RRAT J1913+1330, one of the first discovered RRATs~\cite{mcla06}, was discovered by the Parkes Radio Telescope.
In subsequent study, its period and DM were determined as $\sim$0.923\,s
and 175.64\,$\mathrm{cm^{-3}\,pc}$~\cite{mcla09}, respectively.
Our FAST observations detected bursts in 4 of the 5 epochs.
No burst was seen on MJD 58140.
All \mbox{DMs} in the individual epochs were redetermined from data collected at 400-800\,MHz (the 270-400\,MHz data were discarded due to large scattering in this frequency range). The new values in the de-dispersion process were 175.427, 175.438, 175.389, 175.392\,$\mathrm{cm^{-3}\,pc}$
on MJD 58085, 58144, 58149, and 58179, respectively, slightly lower than the values 
obtained in McLaughlin et al. (2009)~\cite{mcla09} and Losovsky \& Dumsky (2014) \cite{loso14} (175.6\,$\mathrm{cm^{-3}\,pc}$).

From observations with the Parkes Radio Telescope, with a central frequency of 1.4\,GHz
and a bandwidth of 256\,MHz, the burst rate of this source was
estimated to be aboutas approximately 7.3\,hr$^{-1}$~\cite{pall11}. Earlier, McLaughlin et al. (2009) \cite{mcla09} had obtained a burst rate of 13\,hr$^{-1}$
with a central frequency of 1390\,MHz and 256-MHz bandwidth.
Bhattacharyya et al. (2018) \cite{bhat18} re-determined the burst rate as 4.7$\pm$0.2\,hr$^{-1}$.
Our FAST observations yielded an estimated burst rate of 155$\pm$8\,hr$^{-1}$, significantly larger than the previous values.
Figure~\ref{fig6} shows the intensity time series in different epochs in two frequency bands.
The burst activities tended to cluster within short time segments (radiation mode) between periods of no detectable radiation (nulling mode).
The duration of the radiation mode, obtained from the 8 radiation segments, was $3.3\pm1.6$\,min. The nulling modes typically lasted for several tens of minutes.
These segments appeared simultaneously in the same time intervals of the P-band (275$-$775\,MHz) and the L-band (1225$-$1625\,MHz).
This phenomenon largely differs from the radiation activity of RRAT J1854+0306.

\begin{figure*}
  \centering
   \includegraphics[width=0.4\linewidth]{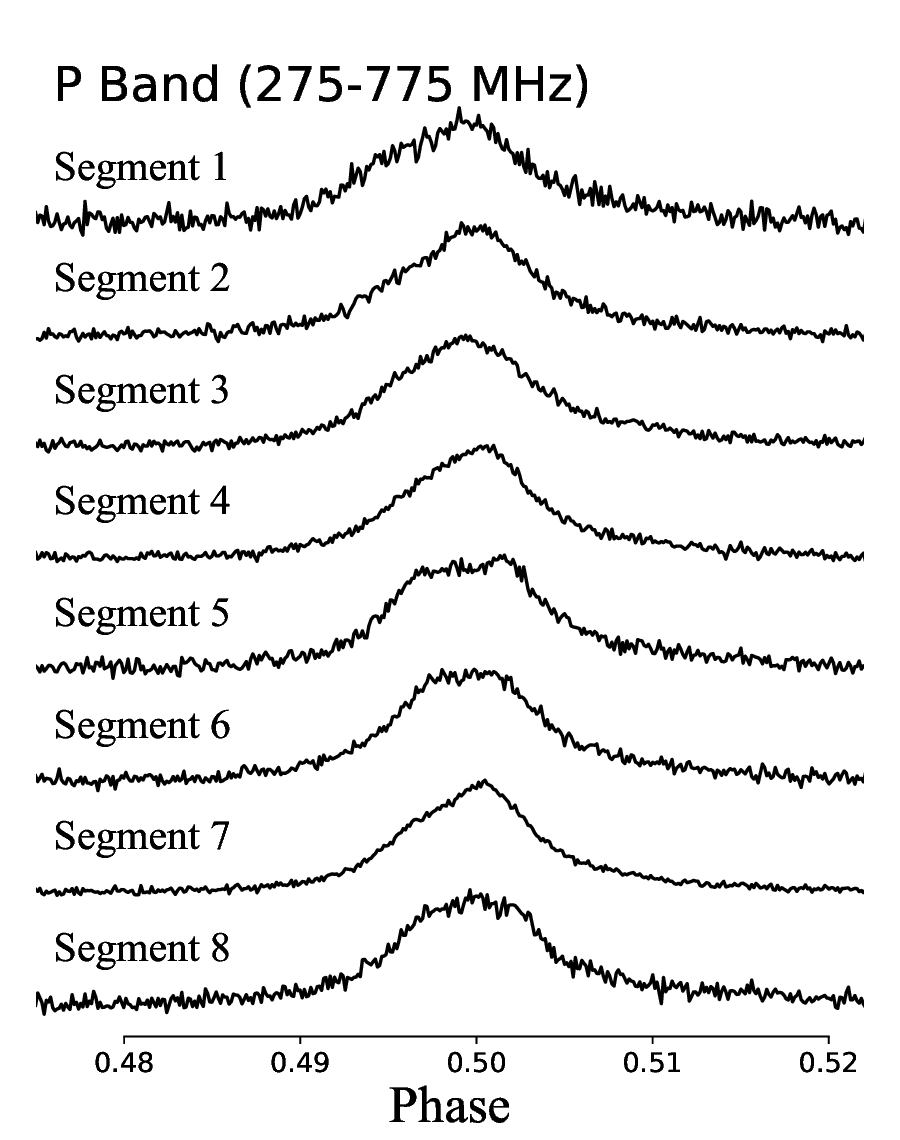}
   \includegraphics[width=0.4\linewidth]{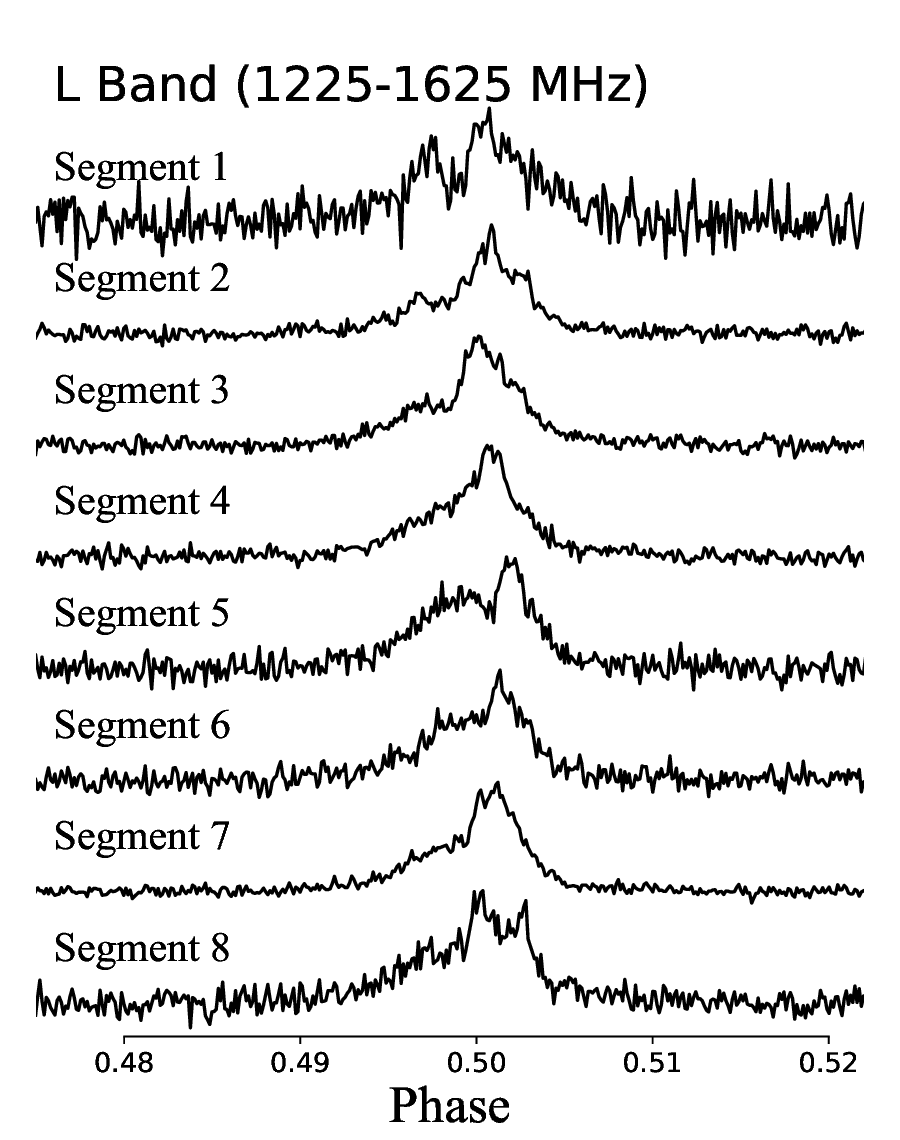}
   \caption{{\small Mean pulse profiles of RRAT J1538+2345 in each radiation segment shown in ~\ref{fig6}.}}
   \label{fig7}
\end{figure*}

\begin{figure}[H]
  \centering
   \includegraphics[width=0.8\linewidth]{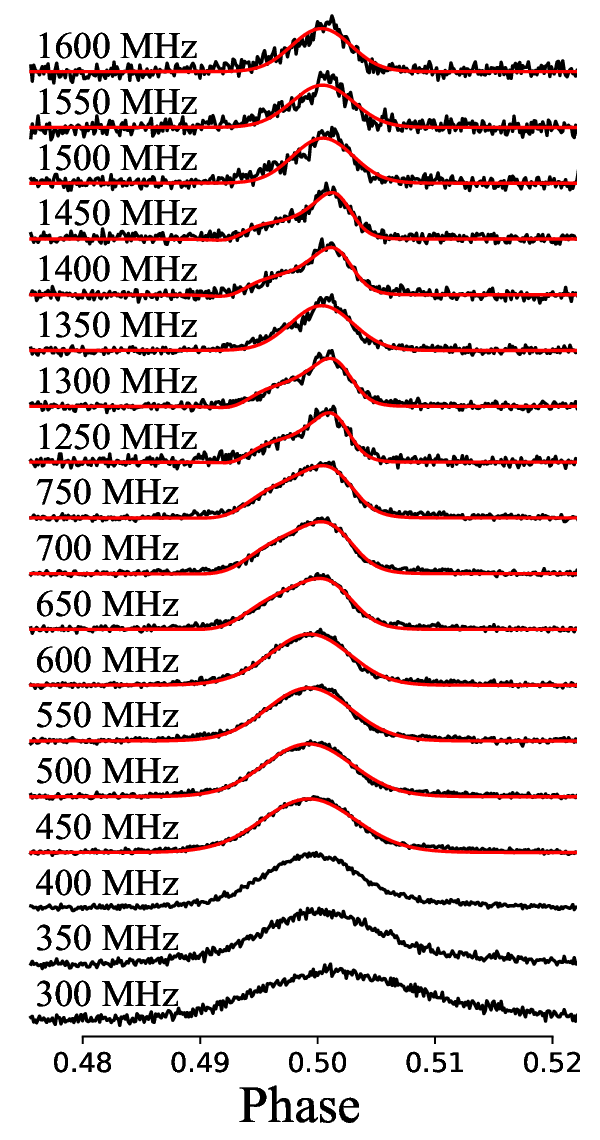}
   \caption{{\small Multi-frequency profiles of RRAT J1913+1330 (black curves).
   The red curves are fitted by Equation~\ref{conalfunction}.
   The narrow bands at 300, 350 and 400\,MHz were not fitted due to the strong scattering widening phenomena 
   in their pulse profiles.}}
   \label{fig8}
\end{figure}

The 8 radiation segments shown in Figure~\ref{fig6} were individually integrated into the mean pulse profiles shown in Figure~\ref{fig7}.
The profiles from different segments were generally consistent, showing a two-component structure
Thus, they can also be treated as not-well-resolved conal-double profile and fitted with Equation~\ref{conalfunction}.
The integrated pulse profiles and fitting curves are plotted as the black and red curves in Figure~\ref{fig8}, respectively.
The narrow bands at frequency 300, 350 and 400\,MHz are not fitted due to the strong scattering
widening phenomena.
With this fitting, the best-estimated inclination angle and impact angle were determined as
$\alpha=42.1\degree$ and $\beta=3.3\degree$, respectively.

\begin{table}[H]
\centering
\footnotesize
\begin{threeparttable}\caption{FRB burst rates obtained from the data of different telescopes.}\label{tab2}
\begin{tabular}{l|cccc}
\toprule
  Source & Telescope & Frequency  & Bandwidth & Burst Rate\\
  &  & (MHz) & (MHz) & (hr$^{-1}$) \\\hline
  \multirow{2}[0]*{RRAT J1538+2345} & LOFAR & 150 & 80 & 66$\pm$7~\cite{kara15} \\
  & GBT & 350 & 100 & 77$\pm$14~\cite{kara15} \\
  & FAST & 1250 & 400 & 302$\pm$12\tnote{1)} \\\hline
\multirow{3}[0]*{RRAT J1854+0306}  & Parkes & 1374 & 288 & 8.9~\cite{kean11}\\
  & Arecibo & 1440 & 100 & 84~\cite{dene09} \\
  & FAST & 1250 & 400 & 102$\pm$10 \\\hline
\multirow{3}[0]*{RRAT J1913$+$1330} & Parkes & 1400 & 256 & 7.3~\cite{pall11}\tnote{2)} \\
  & Parkes & 1390 & 256 & 13~\cite{mcla09} \\
  & Lovell & 1402 & 64 & 1.5~\cite{mcla09} \\
  & Lovell & 1400 & 300 & 4.7$\pm$0.2~\cite{bhat18} \\
  & FAST & 525 & 500 & 155$\pm$8 \\
\bottomrule
\end{tabular}
\begin{tablenotes}
\item[1)] the error calculation assumes Poisson distribution of burst events.
\item[2)] calculated from 138 pulses detected in 13 hours.
\end{tablenotes}
\end{threeparttable}
\end{table}

\section{Discussion}
\label{sect:discussion}

The burst rates obtained in this paper were consistently larger than previous results, suggesting that many of the RRAT burst rate measurements are affected by sensitivity limitations.
They also imply a wide energy distribution of individual bursts, with most of the detections falling in the high-energy end.
Thus, more bursts could potentially be detected by larger telescopes.
In fact, RRATs might be weak pulsars with a giant (or strong) pulse phenomenon,
and the detected bursts might be giant (strong) pulses in the tail of a power-law distribution~\cite{knig06,welt06}.
However, as FAST is in the commissioning phase, accurately determining the system sensitivity and performing other relevant time-domain studies are works in progress, and cannot be commented upon in
the present article.

When modeling the mean pulse profiles, the inclination and impact angles were treated as free
parameters in the fitting procedure.
Nevertheless, it must be pointed out that the fitting results are model dependent and are influenced by
the selected fitting function and initial values.
The conal beam shape was assumed as a square hyperbolic function.
Although Lu et al.~\cite{lu16a} considered this function as the best-fit shape to the pulsar beam,
it might not exactly represent the beam shape.
The initial inclination angle was set to an acute value (45$\degree$),
and the initial impact angle was positive (10$\degree$).
Consequently, an acute inclination angle and a positive impact angle were obtained in the fitting results.
Considering the physical model of pulsar radiation, the acute inclination angle or its
supplementary angle will have the same effect on the above fitting, where the impact angle
could also be a nearly opposite result.
Thus, the fitting may not accurately constrain the radiation geometry.
Nevertheless, the fitting results suggest that the profile is well explained by the conal beam model,
and the radiation beam of RRAT J1538+2345 should contain a core component.

The flux distribution of single pulses from RRAT~J1854+0306 is difficult to obtain, as sufficient burst data are lacking; thus, the pulse profile in each epoch could not be treated
as a typical case.
If these bursts have a power law distribution similar to that of giant pulses,
then strong pulses would dominate the integrated pulse profile.
Alternatively, if the bursts follow the log-normal distribution of normal pulses, $\sim$50 pulses are
almost sufficient for constructing the mean pulse profile.
Thus, despite the small number of pulses in the integration, the difference between the mean profiles of RRAT~J1854+0306 in two epochs might reflect the intrinsic evolution of the pulsar¡¯s magnetosphere.
In reality, Bhattacharyya et al. (2018) \cite{bhat18} pointed out that RRATs have typically larger
period and magnetic field strength, and may have properties similar to magnetars.
Consequently, RRATs may have very active magnetospheres, and their profiles could evolve dramatically over time.
Profile variation, which is observed in radio magnetars, might also appear in some RRATs such as RRAT~J1854+0306. However, this conjecture must be verified in follow up observations over multiple epochs.

As shown in Figure~\ref{fig6}, RRAT J1913+1330 has distinct radiation and nulling modes.
These features are consistent with those reported by Bhattacharyya et al. (2018)~\cite{bhat18}, who reported the typical duration of an active segment as several minutes.
When the data outside the radiation segments in Figure~\ref{fig6} were integrated over each observation,
no pulsations were detected.
RRAT J1538+2345 also showed many segments with continuous pulse trains (Figure~\ref{fig1}), which may be \textbf{similar to the radiation phase in} RRAT J1913+1330.
No noteworthy trend in the flux density evolution emerged in any radiation segment of J1913+1330, or in any continuous pulse train of RRAT J1538+2345.
Therefore, these phenomena might be induced by short-term alterations of the magnetosphere, similarly to the mode-switch phenomenon of pulsars.

Additionally, the ``weak mode'' phenomenon referred in Bhattacharyya et al. (2018)~\cite{bhat18} was absent in our observation.
When the pulses in each nulling segment were integrated into pulse profiles,
no radiation signs were found,
possibly because the duration was limited to $\sim$2.5\,hr,
meaning that the weak emission mode was undetectable.

\section{Summary}
\label{sect:summary}

FAST observed single pulses with high signal-to-noise ratio emitted from three RRATs.
The calculated burst rate of each RRAT was significantly larger than those reported in previous results.
The mean pulse profiles and their temporal evolutions were also investigated.
RRAT J1854+0306 exhibited different mean pulse profiles during two epochs in the 1000$-$1500\,MHz range, which may induce by
variable magnetosphere activities in this source.
RRAT J1913+1330 exhibited apparent radiation segments in its intensity trains,
with simultaneous onsets and ceases in the P-band (275$-$775\,MHz) and the L-band (1225$-$1625\,MHz).
RRAT J1538+2345 showed similar continuous pulse trains in the 1000$-$1500\,MHz frequency range.
This phenomenon is similar to the mode switching phenomenon observed in ordinary pulsars, and may reflect short-term variations of the magnetosphere.
The frequency-dependent evolutions of the mean pulse profiles of RRAT J1538+2345 and RRAT J1913+1330 were also investigated.
The RRAT J1538+2345 and RRAT J1913+1330 profiles at different frequencies were fitted by a cone core model and a conal beam model, respectively.
The inclination and impact angles of both pulsars were estimated from the data fittings.

Interestingly, the significant sensitivity improvement of FAST over existing instruments has already largely ``increased'' the burst rate of RRATs, and revealed more details on their emission activities. Therefore, future RRAT observations with FAST are expected to uncover additional phenomena, providing a more informative classification of RRATs and a better understanding of their emission mechanisms.

%%%%%%%%%%%%%%%%%%%%%%%%%%%%%%%%%%%%%%%%%%%%%%%%%%%%%%%
%%% Acknowledgements. 
%%%%%%%%%%%%%%%%%%%%%%%%%%%%%%%%%%%%%%%%%%%%%%%%%%%%%%%
\Acknowledgements{
This work is supported by the National Key R\&D Program of China under grant number 2018YFA0404703,
the National Natural Science Foundation of China (Grant No. 11225314),
the Open Project Program of the Key Laboratory of FAST, NAOC, Chinese Academy of Sciences,
and the project of Chinese Academy of Science (CAS) and the Max-Planck-Society (MPS) collaboration.
This work made use of the data from the FAST telescope (Five-hundred-meter Aperture
Spherical radio Telescope).
FAST is a Chinese national mega-science facility, built and operated by the
National Astronomical Observatories, Chinese Academy of Sciences.
The FAST FELLOWSHIP is supported by Special Funding for Advanced Users, budgeted and
administrated by Center for Astronomical Mega-Science, Chinese Academy of Sciences (CAMS).
KL acknowledges the financial support by the European Research Council for the ERC Synergy
Grant BlackHoleCam under contract no. 610058.
YLY is supported by by National Key R\&D Program of China (2017YFA0402600) and CAS "Light of West China" Program.
We would like to thank Laura Spitler for carefully proofing read.
}
%%%%%%%%%%%%%%%%%%%%%%%%%%%%%%%%%%%%%%%%%%%%%%%%%%%%%%%
%%% Conflict of interest. ????????????
%%%%%%%%%%%%%%%%%%%%%%%%%%%%%%%%%%%%%%%%%%%%%%%%%%%%%%%
%\InterestConflict{The authors declare that they have no conflict of interest.}

%%%%%%%%%%%%%%%%%%%%%%%%%%%%%%%%%%%%%%%%%%%%%%%%%%%%%%%
%%% Supplements. ????????, ????
%%%%%%%%%%%%%%%%%%%%%%%%%%%%%%%%%%%%%%%%%%%%%%%%%%%%%%%
%\Supplements{}

%%%%%%%%%%%%%%%%%%%%%%%%%%%%%%%%%%%%%%%%%%%%%%%%%%%%%%%
%%% Reference section. 
%%% citation in the content using "some words~\cite{1,2}".
%%% ~ is needed to make the reference number is on the same line with the word before it.
%%%%%%%%%%%%%%%%%%%%%%%%%%%%%%%%%%%%%%%%%%%%%%%%%%%%%%%

\end{multicols}
\end{document}